\documentclass[superscriptaddress,twocolumn,aps,pre,10pt]{revtex4-1}
\pdfoutput=1
\usepackage{times}
\usepackage{amsmath}
\usepackage{amsfonts}
\usepackage{amssymb}
\usepackage{bm}
\usepackage{graphicx}
\usepackage{color}
\usepackage{changes}
\usepackage{float}

\begin{document}
\title{Swarm Hunting and Clusters Turning Inside Out in Chemically Communicating Active Mixtures}

\author{Jens Grauer}
\affiliation{Institute for Theoretical Physics II: Soft Matter, Heinrich-Heine University D\"{u}sseldorf,
            Universit\"{a}tsstra{\ss}e 1, 40225 D\"{u}sseldorf, Germany}
\author{Hartmut L\"{o}wen}
\affiliation{Institute for Theoretical Physics II: Soft Matter, Heinrich-Heine University D\"{u}sseldorf,
            Universit\"{a}tsstra{\ss}e 1, 40225 D\"{u}sseldorf, Germany}
\author{Avraham Be'er}
\affiliation{Zuckerberg Institute for Water Research, The Jacob Blaustein Institutes for Desert Research, Ben-Gurion University of the Negev,
            Sede Boqer Campus 84990,  Midreshet Ben-Gurion, Israel}
\affiliation{Department of Physics, Ben-Gurion University of the Negev,
            Beer Sheva 84105, Israel}
\author{Benno Liebchen}
\email[Electronic mail: ]{liebchen@fkp.tu-darmstadt.de}
\affiliation{Institute for Theoretical Physics II: Soft Matter, Heinrich-Heine University D\"{u}sseldorf,
            Universit\"{a}tsstra{\ss}e 1, 40225 D\"{u}sseldorf, Germany}
\affiliation{Institut f\"ur Festk\"orperphysik, Technische Universit\"at Darmstadt, 64289 Darmstadt, Germany}

\date{\today}

\begin{abstract}
A large variety of microorganisms produce molecules to communicate via complex
signaling mechanisms such as quorum sensing and chemotaxis. The biological
diversity is enormous, but synthetic inanimate colloidal microswimmers mimic
microbiological communication (synthetic chemotaxis) and may be used to explore
collective behaviour beyond the one-species limit in simpler setups. In this work we combine
particle based and continuum simulations as well as linear stability analyses,
and study a physical minimal model of two chemotactic species. We observed a rich phase diagram comprising a
``hunting swarm phase'', where both species self-segregate and form swarms,
pursuing, or hunting each other, and a ``core-shell-cluster phase'', where one
species forms a dense cluster, which is surrounded by a (fluctuating) corona of
particles from the other species. Once formed, these clusters
can dynamically turn inside out, representing a ``species-reversal''. These
results exemplify a physical route to collective behaviours in microorganisms
and active colloids, which are so-far known to occur only for comparatively large and complex animals like insects or crustaceans.

\end{abstract}

\maketitle

Chemotaxis - the movement of organisms in response to a
chemical stimulus - allows them to navigate in complex environments, find food
and avoid repellants.
%
%
It is involved in many biological processes where microorganisms (or
cells) coordinate their motion; these include wound healing, fertilization,
pathogenic invasion of a host, and bacterial colonization
\cite{cejkova:2016,wadhams:2005}. In such cases, microorganisms are attracted
(or repelled) by certain substances (chemoattractants/ chemorepellents), but
they are also attracted to chemicals produced by other microorganisms (or
cells), such as cAMP in the case of Dictyostelium cells \cite{Eidi2017} or
autoinducers in signaling Escherichia coli \cite{Laganenka2016},
which leads to chemical interactions (communication) among the microorganisms.

\begin{figure}[h]
  \centerline{\includegraphics[width=0.53\textwidth]{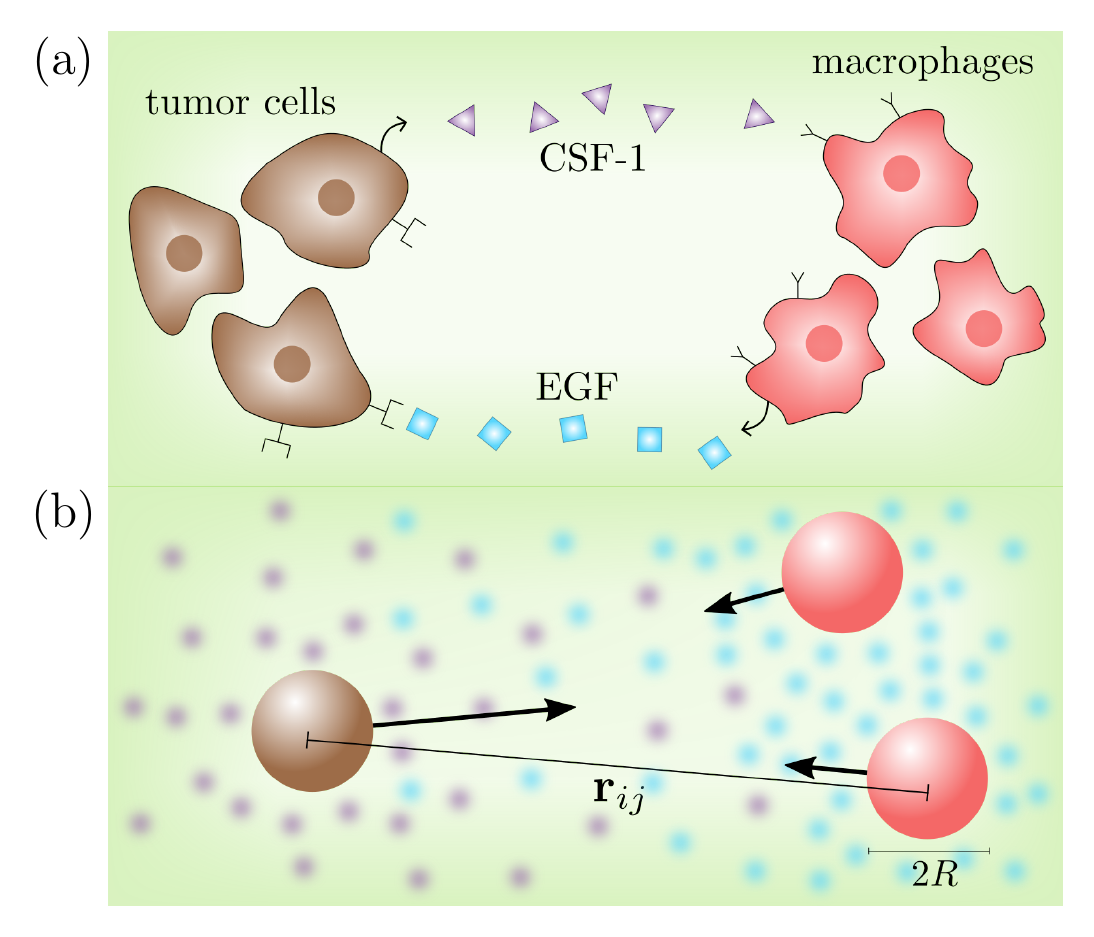}}
  \caption{
  \label{fig:cartoon}
  Schematic: (a) Interaction between tumor cells and macrophages
  (b)~physical minimal model used in the simulation: two species realized as different particles
     (brown and red) with radius $R$ and distance $\mathbf{r}_{ij}$. The movement
     of the particles depends both on their self-produced chemicals (blue and purple) and on the concentration produced by the other species.
     Arrows represent effective chemical interactions among the particles, which in general are non-reciprocal.}
\end{figure}

While many existing models studying microbiological chemotaxis (or chemical interactions) focus on a single species \cite{TindallII,Murray2003,Hillen2008,Painter2018,Painter2011,Dolak2005,Mukherjee2018,Zimmermann2018}, the typical situation
in the microbiological habitat is that various different species simultaneously produce certain
chemicals to which others respond via chemotaxis or based on quorum sensing mechanisms.
One simple example involving chemical signaling across species is provided by macrophage-facilitated breast cancer cell invasion which has recently been modeled \cite{Tumor-Macrophage}. There,
tumor cells attract macrophages, which are certain white blood cells normally playing a key role in the human immune system.
They then control the physiological function of the macrophages and exploit
their abilities. More specifically, the tumor cells produce the
colony-stimulating factor (CSF-1) leading to the attraction and growth of
macrophages which in turn release epidermal growth factors (EGF) resulting in
the growth and mobility increase of the tumor cells (see
Fig.~\ref{fig:cartoon}).

Similarly to microorganisms, synthetic inanimate colloids, coated with a
material which catalyzes a certain reaction on (a part of) their surface, show chemical interactions as well \cite{Stark2018,robertson2018,liebchen_loewen:2018}.
%
There, the colloids act as sources of the chemical field,
which shows a $1/r$-steady-state profile in 3D (if the chemical does not 'decay' e.g. through bulk reactions), leading to long-ranged chemical interactions between the colloids.
For active colloids \cite{Marchetti:2013,Romanczuk2012,Kurzthaler2018,bechinger-dileonardo-loewen-etal:2016,Aranson2013}, these interactions have been explored
in single-species systems
\cite{Saha:2014,Pohl:2014,liebchen2015,liebchen2017,huang2017,liebchen2019}, and
more recently also in mixtures
\cite{soto_golestanian:2014,Schmidt2019,Niu2017,stürmer2019,singh2017,agudo2019,Wang2018},
where chemical interactions can be non-reciprocal and break action-reaction
symmetry \cite{soto_golestanian:2014,Ivlev2015,Sengupta:2011}. This allows for
the formation of active molecules
\cite{soto_golestanian:2014,Schmidt2019,Niu2017}, where self-propulsion
spontaneously emerges when the underlying nonmotile 'colloidal atoms' bind
together. Similarly as for their microbiological counterparts, in all these studies on mixtures of synthetic colloids it has been assumed that the
different species interact via a single chemical substance.


\begin{figure*}
        \begin{center}
    \includegraphics[width=1.\textwidth]{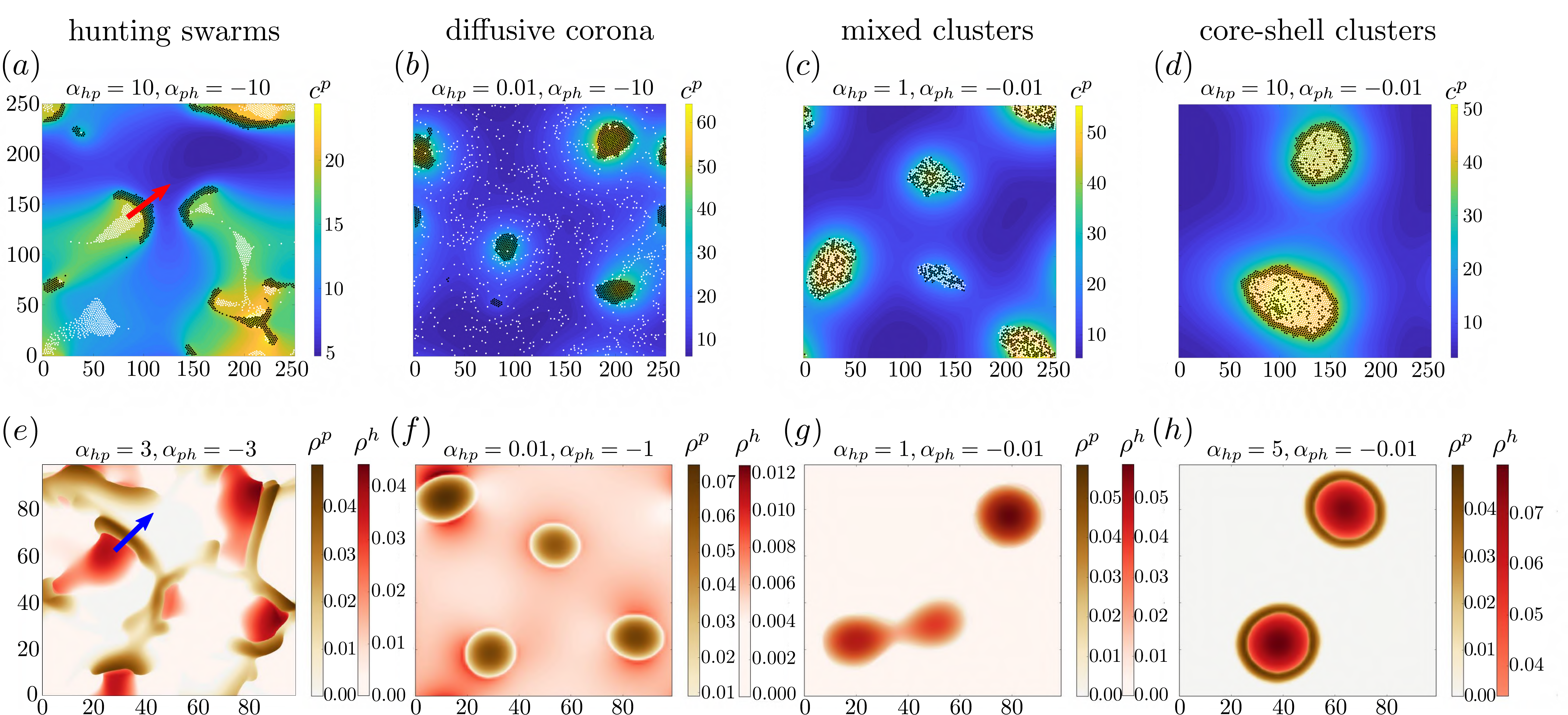}
  \end{center}
  \caption{
  \label{fig:full_snapshots}
  \textit{Hunting swarms} and \textit{core-shell clusters}: Simulation snapshots of Eqs.~(\ref{eq:particles_norm},\ref{eq:phoretic_norm}) for $2N=2000$ chemically interacting particles (white dots represent hunter-particles; black dots show prey-particles) coupled to self-produced chemical fields at time $t=1500$ $(a)$, $5000$ $(b-d)$.
  Panels $(a-d)$ show particle based simulations, where colours show the chemical field produced by the prey $c^p$, panels $(e-h)$ show simulations of the associated continuum equations at time $t=5000$ $(e)$, $10000$ $(f-h)$, where colours show the density of hunters $\rho^h$ and prey $\rho^p$.
  $(a,e)$ show \textit{hunting swarm} patterns, $(c,g)$ show \textit{mixed clusters}, $(b,f)$ show \textit{core-shell clusters} with diffusive and $(d,h)$ with rigid corona.
  Dimensionless parameters (tildes omitted):
  $\alpha_{pp}=1$, $\alpha_{hh}=0$, $\mu=0.001$, $D_c=1$, $D=0.001$ $(a-d)$, $D=0.01$ $(e-h)$ for reasons of stability, $\epsilon=1$ and box length~$L_{\text{box}}=250$ $(a-d)$, $L_{\text{box}}=100$ $(e-h)$.
  See supplementary material for simulation details and the stabilization method used for the field equations underlying panels $(e-h)$. }
\end{figure*}

In the present work,
we propose and explore a physical minimal model for two
species of chemically interacting particles, both of which produce an individual chemical substance. By comparing numerical simulations of Langevin equations describing the particle dynamics (Figs.~\ref{fig:full_snapshots}(a-d))
with numerical solutions of deterministic continuum equations describing the dynamics of their density fields
(Figs.~\ref{fig:full_snapshots}(e-h))
and a linear stability analysis, we systematically explore and analyze the phase diagram of this system.
As our key result, we discover
a ``hunting-swarm phase'' (see Figs.~\ref{fig:full_snapshots}(a,e)), where both species segregate and form individual swarms, one of them closely pursuing the other one.
This phase resembles
a group of hunters chasing a group of prey trying to stay together, not allowing the hunters to split up the group.
The results are important because a similar form of ``swarm hunting'' is known to occur in various systems on larger scales, e.g. in insects and systems of larvae hunting crustaceans (Daphnia) \cite{Boonman2019,jeschke2007,zhdankin2010}
but not for microorganisms or synthetic colloids.
Physically this phase occurs, if one species (``the hunters``) is attracted by the chemicals produced by the other species (''the prey``) and the prey is in turn repelled by the chemicals produced by the hunters.
Hence, the emergence of hunting swarms hinges on the new ingredient of our model - the production of an individual chemical per species.
By systematically exploring the parameter space underlying our model, we find that hunting swarms in fact occur generically if the chemical interactions are strong enough and have opposite sign.
However, if the response of hunters and prey to the chemicals produced by the respective other species is strongly asymmetric, we instead find dense clusters of one species surrounded by a diffusive or rigid corona of particles from the other species (see Figs.~\ref{fig:full_snapshots}(b,d,f,h)).
These core-shell clusters can show a complex dynamics, turning themselves inside out for appropriate initial conditions which is a remarkable process, occuring also in other contexts in nature: The multicellular green alga Volvox, for example, undergoes such an inversion, in which spherical embryos turn their multicellular sheet completely inside out \cite{Goldstein:2015}. The setup considered in the present work allows us to exemplify that a phenomenologically similar reversal may in principle originate from a remarkably simple mechanism  hinging on a systematic invasion of the hunters into a cluster of prey particles, as we will later discuss in detail. The result of this process is a counterintuitive state where the hunters form a dense inner core, surrounded by prey particles which try to stay in close contact to each other.

\section{Model}
We consider an ensemble of two species of overdamped colloids (synthetic or biological), which we call prey and hunters, $s\in \{p,h\}$,
each of which contains $N$ particles which produce a chemical field $c^s({\bf r},t)$ with a rate $k_0$. We assume that each particle responds
via (synthetic) chemotaxis \cite{Saha:2014,liebchen2019} to the gradients of both chemical substances. To model the particle dynamics we use Langevin equations ($i = 1,\ldots,N$, $s\in \{p,h\}$):
\begin{equation} \label{eq:particles}
  \gamma\partial_t \mathbf{r}_i^{s}(t) = \sum_{s' \in \{p,h\}}   \alpha_{ss'} \nabla c^{s'} - {\nabla}_{\mathbf{r}_i} V + \sqrt{2D} \gamma \bm{\eta}_i^{s}
\end{equation}
where $D$ is the translational diffusion coefficient of the particles, $\gamma$ is the Stokes drag coefficient (assumed to be the same for both species) and $\bm{\eta}_i^{s}(t)$ represents unit-variance Gaussian white noise with
zero mean. The chemotactic coupling coefficient of species $s$ to the chemical of species $s'$ is denoted as
$\alpha_{ss'}$ where $\alpha_{s s'} > 0$ leads to
chemoattraction and $\alpha_{s s'} < 0$ results in chemorepulsion (negative chemotaxis). In addition, $V$ accounts for excluded volume interactions among the particles which all have the same radius $R$ and which we model using the Weeks-Chandler-Anderson
potential $V=\frac{1}{2} \sum_{i,j \neq i} V_{ij}$ where the sums run over all particles and where
$V_{ij} = 4\epsilon \left[ (\frac{\sigma}{r_{ij}})^{12} - (\frac{\sigma}{r_{ij}})^6 \right] + \epsilon$ if $r_{ij} \leq 2^{1/6}\sigma$ and zero else.
%
%
Here $\epsilon$ determines the strength of the potential, $r_{ij}$ denotes the
distance between particles $i$ and $j$, $r_c=2^{1/6} \sigma$ indicates a cutoff radius beyond which the potential energy is zero and $\sigma=2R$ is the particle diameter.

The chemical fields $c^h(\mathbf{r},t),c^p(\mathbf{r},t)$ are produced by particles of hunters and prey, respectively. The dynamics of these fields, follows a diffusion equation (diffusion coefficient $D_c$), with
additional (point) sources. We also use a
sink term whose coefficient may be zero or nonzero if chemical reactions or other processes degrading the chemical occur in bulk.
For simplicity we focus on the case where $D_c,k_0,k_d$ are identical for both species.
\begin{eqnarray} \label{eq:phoretic}
  \partial_t c^{s}(\mathbf{r},t) &=& D_c \Delta c^{s} - k_d c^{s} + k_0 \sum_{i=1}^N\delta(\mathbf{r} - \mathbf{r}_i^{s}) \; . \;\;
\end{eqnarray}
To reduce the parameter space, we choose $x_0=R$ and $t_0=\frac{1}{k_0}$ as the units of length and time.
The resulting dimensionless parameters are
$\tilde{\alpha}_{kl} = \frac{\alpha_{kl}}{\gamma k_0 R^{d+2}}$, $\tilde{\epsilon}=\frac{\epsilon}{\gamma
k_0 R^2}$ $\tilde{D_c} =
\frac{D_c}{k_0 R^2}$, $\tilde{D}=\frac{D}{k_0 R^{2}}$ and $\mu=\frac{k_d}{k_0}$
(see the Supplementary Material (SI) for details) and equations
(\ref{eq:particles},\ref{eq:phoretic}) reduce to (omitting tildes)
\begin{eqnarray}
  \label{eq:particles_norm}
 &&\partial_t \mathbf{r}_i^{s}(t) = \sum_{s \in \{p,h\}}   \alpha_{ss'} \nabla c^{s'} - {\nabla}_{\mathbf{r}_i} V + \sqrt{2D} \bm{\eta}_i^{s} \\
  \label{eq:phoretic_norm}
  &&\partial_t c^{s}(\mathbf{r},t) = \left( D_c \Delta
   - \mu \right) c^{s} + \sum_{i=1}^N\delta(\mathbf{r} - \mathbf{r}_i^{s})
\end{eqnarray}

\section{Hunting Swarms and Core-Shell Clusters}
To explore the collective behaviour of many chemotactic agents, we now solve equations (\ref{eq:particles_norm}) and (\ref{eq:phoretic_norm}) using Brownian dynamics simulations for the particle dynamics coupled to a finite difference scheme to calculate the dynamics of the self-produced chemical fields. We solve the diffusion equation in 2D for numerical efficiency and do not expect that our results would change qualitatively when solving the 3D diffusion equation (see the exemplaric simulation snapshot Fig.~1 in the SI and notice that the linear stability analysis which does not depend on the dimensionality of the diffusion equation is also in very good agreement with the particle based simulations). We use a quadratic simulation box with periodic boundary conditions (see SI for details) and
observe the following patterns or nonequilibrium phases:
\begin{enumerate}
  \item[i)] a hunting swarm phase (see Figs.~\ref{fig:full_snapshots}(a,e) and movies 1,5), where both species
  segregate and form moving swarms which hunt each other
  \item[ii)] a clustering phase (see Figs.~\ref{fig:full_snapshots}(c,g) and movies 3,7), where both species form a cluster and the different species are mixed
  \item[iii,iv)] two phases showing core-shell clustering, where one species forms the inner core and the
  other one forms a corona which may be diffusive (b,f) or rigid and which is strongly localized around the core (d,h).
\end{enumerate}

\begin{figure*}
  \begin{center}
    \hspace{-1cm}
    \includegraphics[width=1\textwidth]{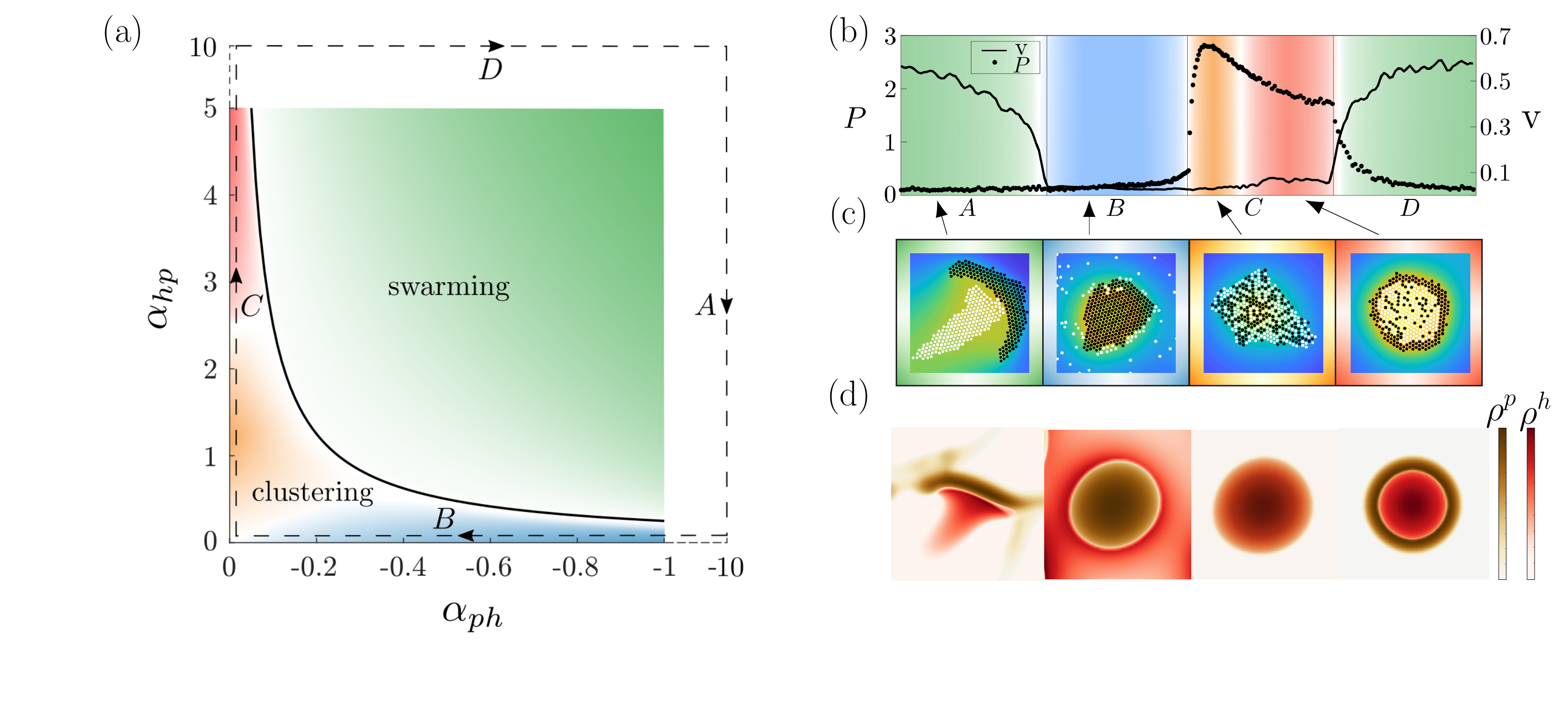}
  \end{center}
  \vspace{-0.2cm}
  \caption{
  \label{fig:phase_plot}
  (a) State diagram in the plane spanned by the chemotactic cross coupling coefficients $\alpha_{ph}$ and $\alpha_{hp}$ for
  fixed $\alpha_{pp}=1$; $\alpha_{hh}=0$. The green domain represents hunting swarms, which are characterized by their ballistic motion and their emergence from an oscillatory instability (the black line shows the analytical prediction of the transition line), whereas colours for the remaining cluster phases are defined via the
  value of the mixing parameter shown in panel (b) (see text). The state diagram was created with more than $200$ evaluated state points.
  (b) Mixing parameter $P$, counting the average number of black next neighbors per white particle and mean particle velocity $v$ at late times
  discriminating between the individual states: Each point corresponds to a parameter set on the dashed line in the parameter plane of panel a.
  The labels $A,B,C,D$ correspond to those shown on the dashed line in panel (a).
  (c,d) Extracts from the simulations underlying Fig.~\ref{fig:full_snapshots} (see movies 1-8).
  }
\end{figure*}

Let us now characterize these phases and the dynamics leading to their emergence in detail.

To see in which parameter regimes each of these patterns prevails, in
Fig.~\ref{fig:phase_plot} we show a slice through the state diagram in the plane of the chemotactic cross-species coupling coefficients
$\alpha_{ph}<0$ and $\alpha_{hp}>0$.
Here we fix $\alpha_{pp}=1$ and $\alpha_{hh}=0$ so that prey-particles chemo-attract each other whereas
the hunter-particles
do not, but note that the specific values choosen here do not have much impact on the emerging patterns.

\textbf{Hunting Swarms:}
The green area in Figs.~\ref{fig:phase_plot}(a,b,c) (movies 1,5) represents the hunting-swarm phase which generically occurs if
$|\alpha_{hp} \alpha_{ph}|$ is large enough, as we will later show using a linear stability analysis.
Here the chemicals produced by the black-coloured particles in Fig.~\ref{fig:full_snapshots}(a) (``prey'') attract the white coloured particles (``hunters''), whereas the
hunter-produced chemicals repel the prey. This results in a swarm of ``prey''
pursued by a swarm of ``hunters''. When two or more prey-swarms collide, the pursuing hunters produce a ``cage'' of high chemical density repelling the prey and trapping it temporarily in a small spatial domain.
The prey then 'evades' sidewards to escape from the hunter-fronts, forming new swarms moving perpendicular to the original ones (see movies 1,5).

\textbf{Core-shell clusters:}
When decreasing $\alpha_{hp}$ (blue domain in Figs.~\ref{fig:phase_plot}(a,b) and movies 2,6),
so that the prey chemo-attracts the hunters only weakly,
we observe that the prey aggregates and forms dense clusters, surrounded by a diffusive corona of hunters.
Surprisingly, when staying with a large
$\alpha_{hp}$ but decreasing $\alpha_{ph}$ instead (red domain in Figs.~\ref{fig:phase_plot}(a,b,c)),
so that the hunters are strongly chemo-attracted by the prey, but the prey has only a weak tendency to avoid the hunter-produced chemicals, we see the opposite case: Although not attracting each other, the hunter-particles form a dense core, surrounded by the prey-particles (red domain in Fig.~\ref{fig:phase_plot}(c) and right panel of Fig.~\ref{fig:phase_plot}(d) and movies 4,8).
To see how these remarkable clusters emerge, let us explore the dynamics underlying their formation.
Initially, the prey-particles, which chemo-attract each other aggregate and form very small clusters.
This leads to an enhanced chemical production, also attracting the hunters,
which subsequently invade the cluster, because $|\alpha_{hp}|>|\alpha_{pp}|$.
Consequently, as more and more hunters enter the cluster, the density of $c^h$ increases in the cluster center, repelling the prey. Since the prey-particles, in turn, couple stronger to their self-produced chemicals than to those produced by the hunters $|\alpha_{pp}|>|\alpha_{ph}|$, they do not flee from the cluster but try to stay together. While in the simulations underlying Fig.~\ref{fig:full_snapshots}, the hunters invade even small prey-clusters, for appropriate initial conditions, we can see a proper inside out reversal of comparatively large clusters (movie 9) (species reversal). In each case, the result is a counterintuitive pattern consisting of a dense cluster containing mostly hunters surrounded by ring of prey-particles.

\textbf{Irregular aggregation:}
Finally, when $\alpha_{ph}$, $\alpha_{hp}$ are both small, with $|\alpha_{ph}|<|\alpha_{hp}|$ (orange regime in Figs.~\ref{fig:phase_plot}(a,b) and movies 3,7), prey and hunter particles form clusters containing a seemingly irregular mixture of hunter and prey particles (Fig.~\ref{fig:phase_plot}(c), orange). These clusters emerge because we have a chemically mediated prey-prey attraction and a hunter-by-prey attraction which exceeds the prey-by-hunter-repulsion, so that effectively prey particles similarly strongly attract all other particles, leading to a rather irregular aggregation.

\textbf{Classification:}
In contrast to the static clusters, structures in the green region of Fig.~\ref{fig:phase_plot}(a) move ballistically and hence show a non-vanishing velocity. Figure \ref{fig:phase_plot}(b) depicts the mean particle velocity $v(t)$ (see SI for details) at late times for parameters chosen along the dashed line in Fig.~\ref{fig:phase_plot}(a), where one can easily see how the velocity in regions of hunting swarms exceeds that in other regimes.
While the hunting swarm phase, which emerges from an oscillatory instability, as discussed further below, can be clearly distinguished from the stationary cluster phases, let us define
an ``order parameter'' $P$ to distinguish the remaining cluster phases.
We define $P$ as the average number of black next neighbors (prey) per white particle (hunter), where we denote a neighbor as a particle within a distance $r_{ij}<2+0.1$.
Figure \ref{fig:phase_plot}(b) shows
$P$ for parameters chosen along the dashed line in Fig.~\ref{fig:phase_plot}(a).
This parameter would have a value of $3$ for completely irregular and infinitely large dense clusters. For the orange domain, where particles aggregate almost irregularly, it has a value $P>2.5$, whereas red means ($1.5<P<2.3$)
and blue means $P<0.5$. Crossover regions between the individual patterns are marked by white domains in Fig.~\ref{fig:phase_plot}(a).

\section{Linear stability analysis -- emergence and dynamics of patterns at early times}
To understand the structure of the state diagram we now introduce a
continuum description for the particle dynamics and perform a linear stability analysis.

\textbf{Continuum model:}
The Smoluchowski equation, describing the dynamics of the (non-normalized) probability $\rho^{s}(\mathbf{r},t)$ to find a particle of species $s$ at position ${\bf r}$ at time $t$ reads as follows ($s \in \{p,h\}$):
\begin{equation}
  \partial_t \rho^{s} = D \Delta \rho^{s} - \sum_{s' \in \{p,h\}} \alpha_{ss'} \nabla \cdot (\rho^{s} \nabla c^{s'}) \;\; , \;\;\;
  \label{eq:partial_rho}
\end{equation}
These deterministic equations are equivalent
to the Langevin equations (\ref{eq:particles}) for point particles ($V=0$). We can now also rewrite the evolution equation for the chemical fields as follows:
\begin{eqnarray}
  \partial_t c^{s} = \left( D_c \Delta -\mu \right) c^{s} + \rho^{s}  \; .
  \label{eq:partial_c}
\end{eqnarray}
Before carrying out a linear stability analysis, let us solve these equations numerically to test them: Integrating Eqs.~(\ref{eq:partial_rho}, \ref{eq:partial_c})
for a uniform initial state (plus small fluctuations) on a square box of size $L_{box}=100$, we indeed find the same patterns as in our particle based simulations (Figs.~\ref{fig:full_snapshots}(e-h) and Fig.~\ref{fig:phase_plot}) (see SI for details regarding these simulations and the used method to stabilize them).

\textbf{Linear stability analysis:}
We now linearize these four coupled equations around the
stationary solution $(\rho,c)=(\rho_0,\rho_0/\mu)$, which represents the uniform disordered phase, and solve them in Fourier Space, to understand the dynamics of a small plane wave perturbation with wavenumber $q$ around the uniform phase.
We denote the dispersion relation of these fluctuations as $\lambda(q)$. If $\lambda$ has a positive real part for some $q$ value, the uniform phase is unstable.
Calculating $\lambda$ (see Supplementary Material for details), we find that the uniform phase looses stability if
\begin{eqnarray}
  2D \mu < \rho_0 Re \left[ \alpha_{pp}+\sqrt{4\alpha_{ph} \alpha_{hp} + \alpha_{pp}^2} \right] \; .
  \label{eq:Keller-Segel-inst}
\end{eqnarray}
where we have choosen $\alpha_{hh}=0$ as in our simulations.
This criterion is consistent with all simulations shown in Fig.~\ref{fig:phase_plot} and
quantitatively agrees with additional simulations which we have performed (not shown). The instability criterion shows that chemo-attractions among the
prey particles support the emergence of a pattern in competition with diffusion and the potential decay of the chemical, whereas cross interactions only support the emergence of a pattern if they, $\alpha_{ph}$ and $\alpha_{hp}$, have the same sign.

To understand the transition between static clusters and hunting swarms,
we also derive a criterion discriminating between stationary instability (static clusters, $\lambda$ is real) and oscillatory instabilities (moving structures, complex $\lambda$) which
reads as follows (see SI):
\begin{equation}
2\sqrt{-\alpha_{ph} \alpha_{hp}} > \alpha_{pp} \; .
\label{eq:inst_imag}
\end{equation}
This criterion defines the solid black line in Fig.~\ref{fig:phase_plot}(a), which quantitatively agrees with our simulations.
It shows that an oscillatory instability and hence moving patterns can appear only if $\alpha_{ph}$,$\alpha_{hp}$ have opposite sign, i.e. if one species effectively hunts the other one, whereas the other one tries to escape.

\begin{figure}
  \centerline{\includegraphics[width=0.5\textwidth]{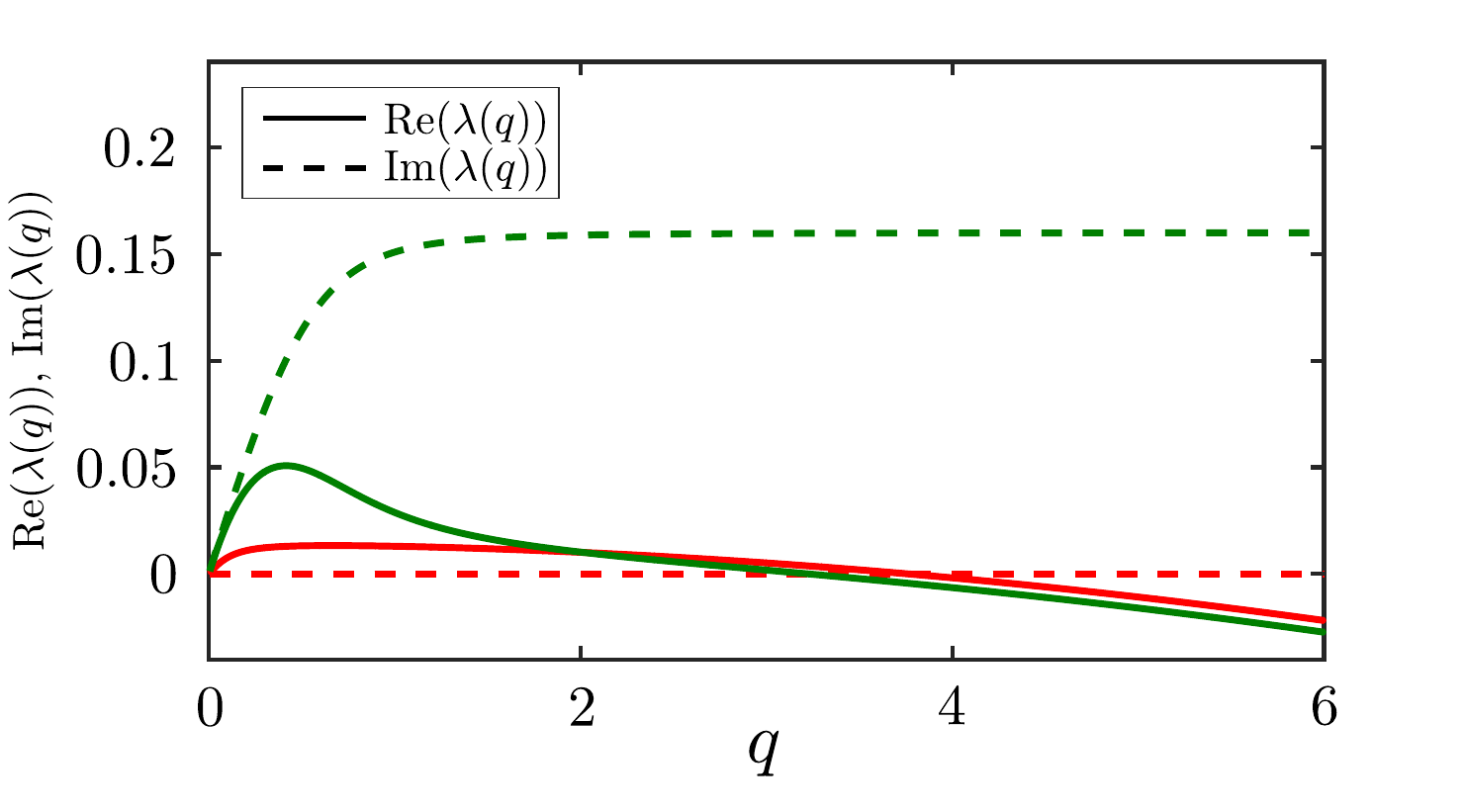}}
  \caption{
  \label{fig:lambda_plot}
  Real and imaginary part of the dispersion relation $\lambda$, for hunting swarms (green)
  and static clusters (red).
  Parameters as in Figs.~\ref{fig:full_snapshots}(a) and (d).}
\end{figure}

In Fig.~\ref{fig:lambda_plot} we show the complete dispersion relation $\lambda(q)$ (real and imaginary part) of small plane wave fluctuations around the uniform phase. Here the location of the maxima in $\mathrm{Re}[\lambda(q)]>0$ define the fastest growing mode, typically determining
the length scale of the pattern at early times.

\begin{figure}
  \centerline{\includegraphics[width=0.5\textwidth]{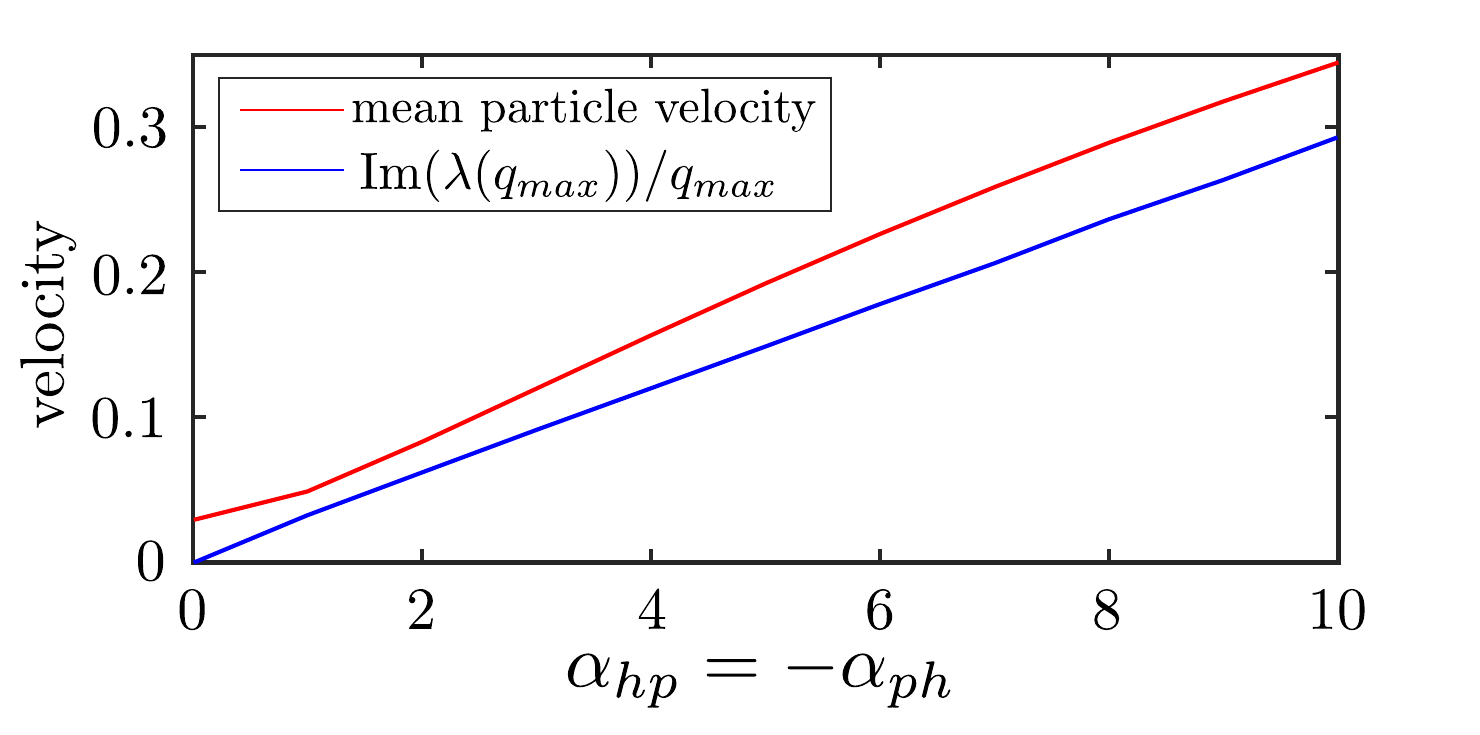}}
  \caption{
  \label{fig:speed}
  Mean particle velocity in the hunting swarm phase, extracted from the simulations underlying Fig.~\ref{fig:phase_plot}(a) at early times (red) and reduced imaginary part of $\lambda$ at the wavenumber
  corresponding to the fastest growing mode, i.e. $\mathrm{Im}(\lambda(q_{max}))/q_{max}$ (blue) as a function of
  $-a_{ph}=a_{hp}$.}
\end{figure}

Having understood the transition line between the cluster phases and the hunting swarms, let us also explore if we can understand how fast the swarms move. To do this, in Fig.~\ref{fig:speed}, we compare the
imaginary part of $\lambda$ (the expected speed of the hunting swarm is $v(q)=\frac{\mathrm{Im}(\lambda)}{q} \rightarrow \langle v \rangle \approx \frac{\mathrm{Im} \lambda(q_{max})}{q_{max}}$) with the velocity of the hunting swarms in our particle based simulations at early times and find close agreement.

%

\section{Structure and growth at late times}
\begin{figure}
  \centerline{\includegraphics[width=0.5\textwidth]{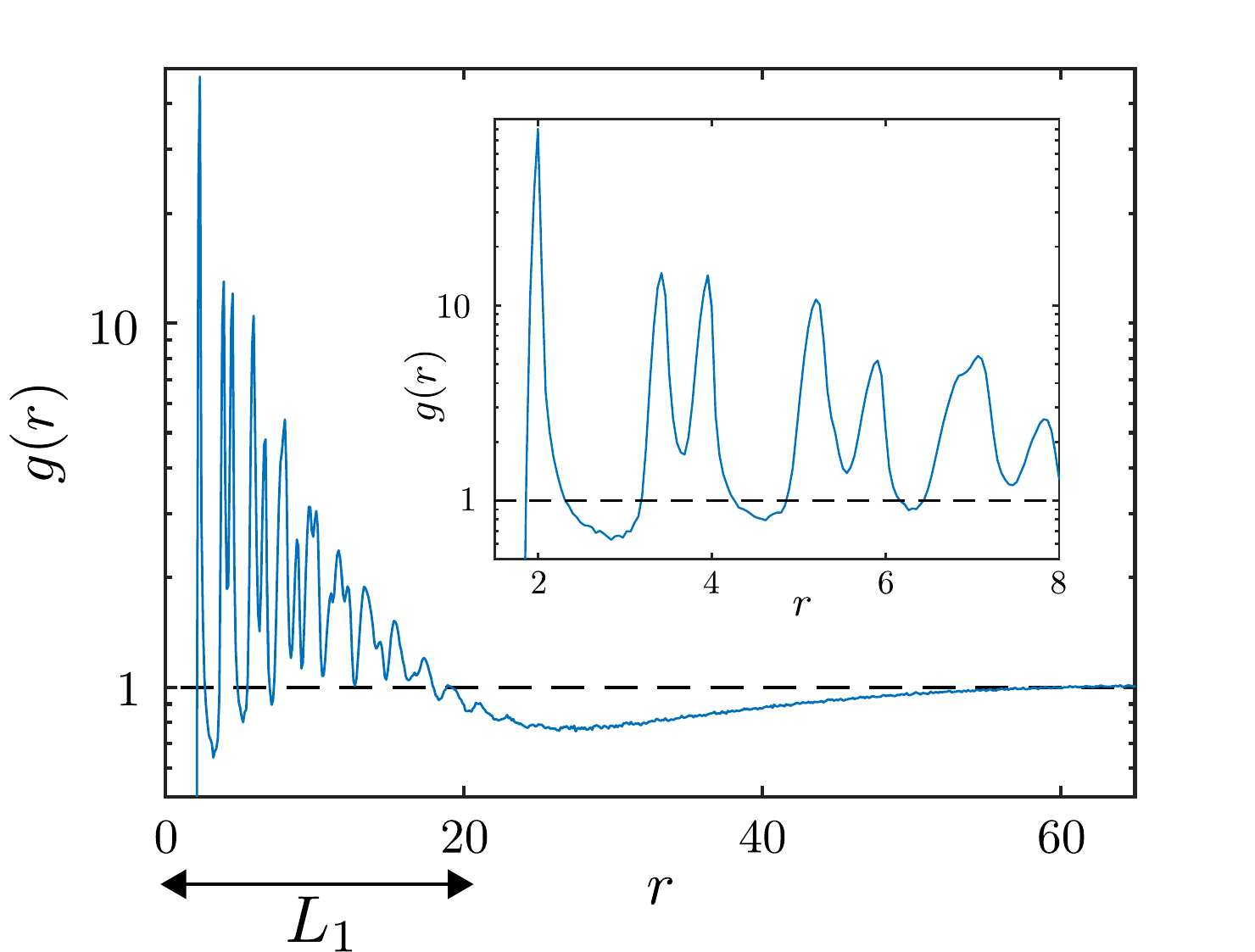}}
  \caption{
  \label{fig:gr}
  Pair-correlation function $g(r)$ (radial average of $g(\mathbf{r})$) of a system of $2N=2000$ particles at time
  $t=250$. The data are averaged over $100$ independent ensembles. The dashed line shows a threshold to extract a characteristic length scale.
  Parameters as in Fig.~\ref{fig:full_snapshots}(d).}
\end{figure}

Having explored how the patterns emerge and behave at early times, we now want to explore their structure and
dynamics also at late times.
To do this, we introduce the instantaneous
pair-correlation function $g(\mathbf{r})$ defined as
\begin{equation}
g(\mathbf{r}) = \frac{1}{\rho_{\text{id}}} \left\langle \sum_{i \ne 0} \delta(\mathbf{r} - \mathbf{r}_i) \right\rangle,
\label{eq:g_r}
\end{equation}
for an average number density $\rho_{\text{id}}=\frac{2N}{L_{\text{box}}^2}$ with
box length~$L_{\text{box}}=250$, total number of particles $2N$ and $\langle \cdot
\rangle$ denoting the ensemble average. The radially averaged and time averaged pair-correlation function $g(r)$, where $r=|\mathbf{r}|$,
shown in Fig.~\ref{fig:gr} describes how the density varies as a function of
distance from a reference particle at which we averaged over all particles of hunters and prey.

As one can see in the inset of Fig.~\ref{fig:gr}, there is a large
peak around $r=2$, which is the typical distance between two particles
($R=1$ in dimensionless units). We can also find peaks around $r=2 \sqrt3$ and
$r=4$ caused by the next two neighbors. This reflects the fact that the
static clusters (blue, orange, red) show a hexagonal packing.

\textbf{Late-stage dynamics:}
Once the patterns have emerged, they reach a state where their morphology changes only slowly. However, even at late stages the size of the individual structures still increases in time, partly due to diffusive processes (coarsening), partly due to ``collisions`` of different clusters (coalescence).

\begin{figure}
  \centerline{\includegraphics[width=0.55\textwidth]{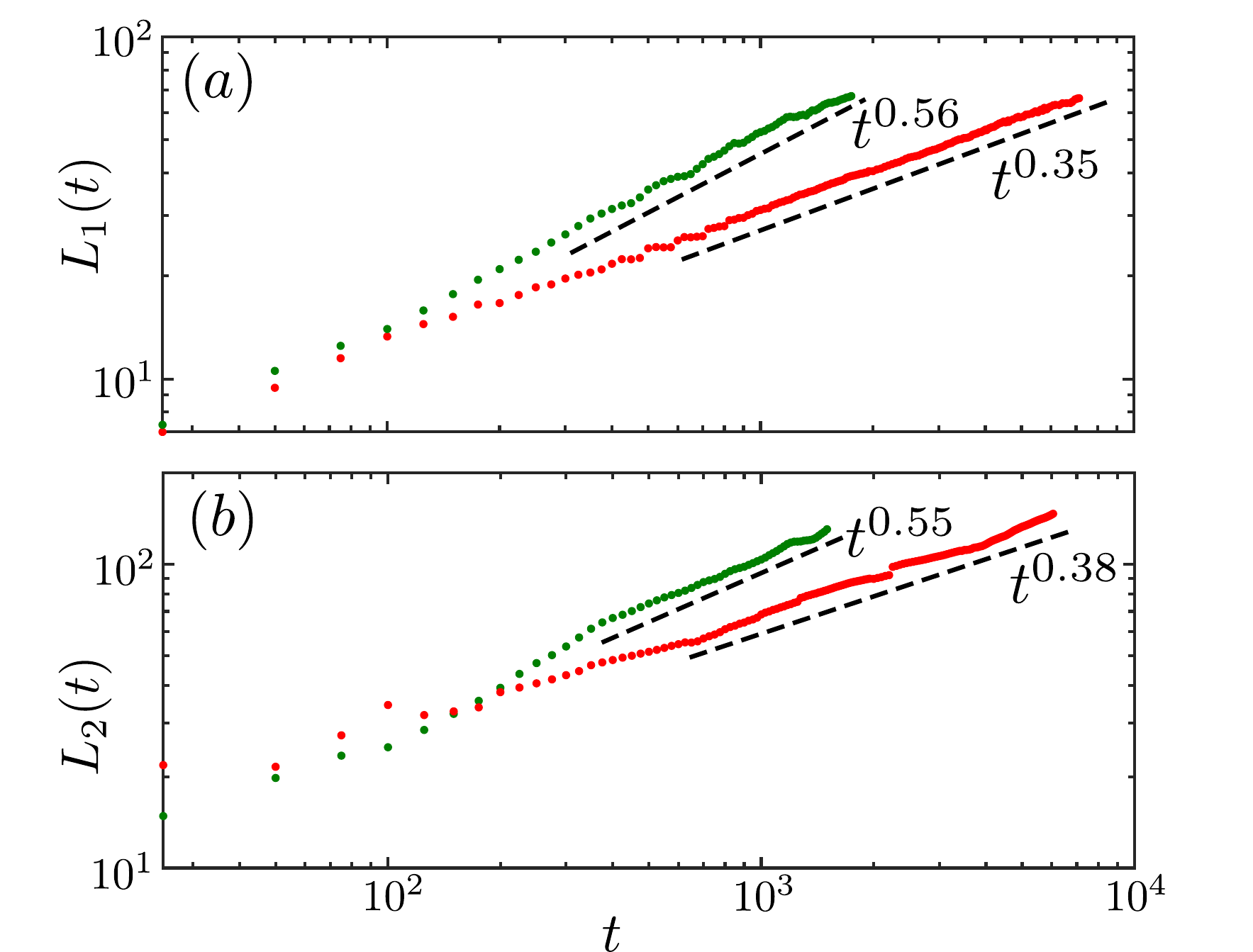}}
  \caption{
  \label{fig:char_length}
  Time-dependent characteristic length scale (a) obtained from the
  pair-correlation function and (b) from the structure factor for structures in the red region (red dotted line) and
  in the green region (green dotted line of Fig.~\ref{fig:phase_plot}(a)).
  The dashed lines indicate the fitted
  exponents.
  Parameters as in Figs.~\ref{fig:full_snapshots}(a,d).}
\end{figure}

To quantify this growth, we consider the time evolution of the radial
distribution function $g(r,t)$ and define the
length scale $L_1(t)$ of clusters as the smallest value
where $g(r,t) \leq 1$, for all $r>L_1$. Thus, the $g(r)$ shown in Fig.~\ref{fig:gr}
corresponds to a length scale of $L_1\approx 20$ (dimensionless units). At late-times, we find that $L_1(t) \propto t^{\beta}$
follows a power law with an exponent of $\beta \approx 0.35$ (Fig.~\ref{fig:char_length}) for the (nonmoving) cluster phases, which is close to the value of
$\beta=\frac{1}{3}$ as expected for diffusive growth (in the absence of hydrodynamic interactions)
\cite{lifshitz-slyozov:1961,bray:2002,gonnella:2015,laradji:2005,camley:2011}.
We find a much larger exponent, of
$\beta \approx 0.56$ (Fig.~\ref{fig:char_length}), for the patterns in the green region, which is a consequence of the fact that the individual structures move ballistically, collide and merge with each other much faster (but also break up).

\begin{figure}
  \centerline{\includegraphics[width=0.5\textwidth]{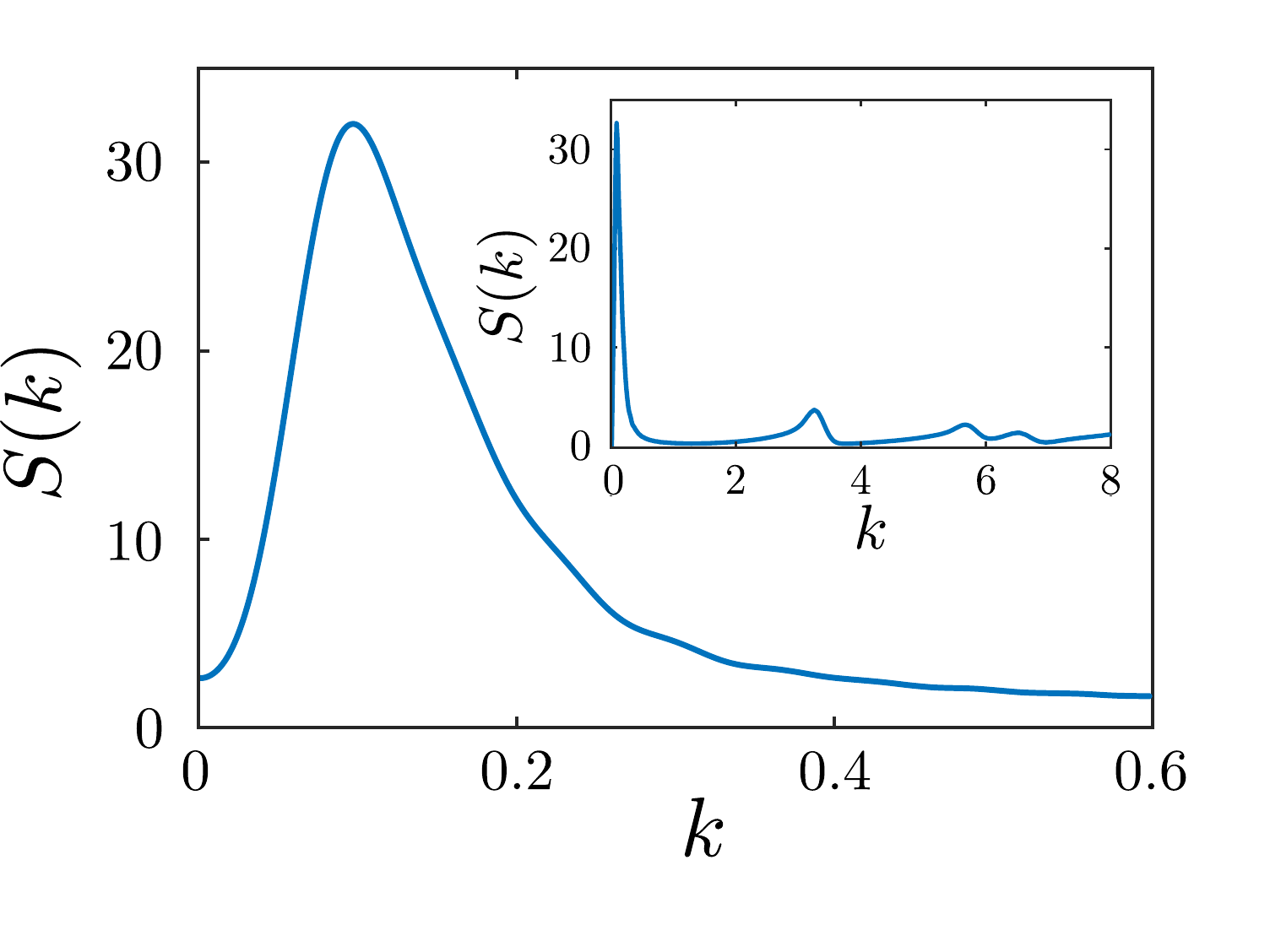}}
  \caption{
  \label{fig:s_k}
  Small wavenumber regime of the structure factor $S(k)$ for a system of $2N=2000$ particles at time
  $t=250$. Inset shows $S(k)$ for a larger wavenumber regime. The wavenumber $k$ is given in dimensionless units ($k_0=\frac{2\pi}{R}$).
  Parameters as in Fig.~\ref{fig:full_snapshots}(d).}

\end{figure}

As a second measure for the growth of the clusters,
we measure the distance between them. To do this,
we consider the structure factor of the system:
\begin{equation}
S(\mathbf{k}) = 1 + \rho_{id} \int_{V} \mathrm{d} \mathbf{r} e^{-i \mathbf{k} \cdot \mathbf{r}}  [g(\mathbf{r})-1] \; .
\label{eq:structure factor}
\end{equation}
%
%
and calculate the distance between clusters as the
inverse of the first moment of the structure
factor \cite{stenhammar:2013}, i.e. as:
\begin{equation}
L_2(t) = 2 \pi \left[ \frac{\int_{2 \pi / L}^{k_{cut}} k S(k,t) \mathrm{d}k}{\int_{2 \pi / L}^{k_{cut}} S(k,t) \mathrm{d}k} \right]^{-1} \; ,
\label{eq:first_moment_s}
\end{equation}
where we choose the cutoff wavelength $k_{cut}$ as the first local minimum of
$S(k)$ \cite{stenhammar:2013}. Figure ~\ref{fig:s_k} shows the structure factor
for a cluster in the red region of Fig.~\ref{fig:phase_plot}(a) at time $t=250$
for small values of $k$. The peaks that can be seen in the inset of
Fig.~\ref{fig:s_k} correspond to the distance of two possible lattice planes of
the hexagonal structure. The peak at $k=3.3$ results from the minimum distance
between two particles ($\frac{2 \pi}{3.3} \approx 2$). One finds a huge peak
around $k=0.11$ with which we can estimate  a typical length, $l \approx \frac{2
\pi}{k} = 57.1$; the enormous size of the peak hinges on the fact that each of
the contributing clusters contains a large number of particles. The $k$-value
where this peak occurs corresponds to the mean cluster distance, which
corresponds to the value of $r$ where $g(r)$ approaches $1$ from below (see
Fig.~\ref{fig:gr}). This distance grows basically with the same power law as the
cluster sizes, as shown in Fig.~\ref{fig:char_length}, i.e. calculating cluster
sizes via $L_1(t)$ and calculating cluster-distances $L_2(t)$ basically leads to
the same growth law (Fig.~\ref{fig:char_length}) \cite{STANICH2013444}. Thus,
there is only one independent macroscopic length scale in the system.

\section{Conclusions}

Inspired by the generic presence of multi-species chemotaxis in microbiological communities, e.g. in
macrophage-tumor cell systems, we have proposed and explored a physical minimal model to study the collective behaviour beyond the commonly considered one-species limit. We have found that the novel key ingredient of our model - the species selective chemical production - leads to patterns which have so-far been known only to occur within the animal kingdom at much larger scales, e.g. in insects and larvae-crustacean-systems: these patterns comprise a ``hunting swarm'' phase consisting of a crowd of
particles of one species pursuing the other species, and a phase where the two-species self-aggregate in a core-shell structure, which may feature a dynamical inside-out reversal en route to the steady state.

All these patterns could be observed both on the level of a particle-based
description (Eqs.~(\ref{eq:particles_norm}, \ref{eq:phoretic_norm}) and in a
continuum model (Eqs.~(\ref{eq:partial_rho}, \ref{eq:partial_c})), allowing to
analytically understand the transition line between cluster phases, which
originate from a stationary instability of the uniform phase, and hunting
swarms, emerging from an oscillatory instability.
As a further characteristic difference between these phases, we find that
clusters (and the distance between them) grow diffusively ($L(t) \propto t^{0.35}$ \cite{STANICH2013444,lifshitz-slyozov:1961,bray:2002,gonnella:2015,laradji:2005,camley:2011}),
whereas hunting swarms grow significantly faster ($L(t) \propto t^{0.56}$ \cite{Cremer2014}).


%

Future work might include more specific biological details and could address the effect of confining boundaries or obstacles \cite{Morin2016,Toner2018,huang2017_2,Rahmani2019}. Other topics concern additional aligning interactions
and their impact on the cluster structure \cite{Das2017,Mones2015,Nilsson2017} and ternary systems describing species of
a longer biological food chain.

\section*{Data availability}
All relevant data are available from the authors upon reasonable request.

\bibliography{paper}

\section*{Author contributions}
H.L. and B.L. have planned and designed the research project; J.G. has performed most of the research; all authors have discussed the results; J.G. and B.L. have written the manuscript; H.L. and A.B. have edited it.

\begin{acknowledgments}
H.L. was supported by the German Research Foundation (DFG) within the SPP 1726 (project LO 418/17-2).
A.B. thanks partial support from The Israel Science Foundation Grant 373/16.
\end{acknowledgments}

\end{document}


\title{Supplemental Material - Swarm Hunting and Clusters Turning Inside Out in Chemically Communicating Active Mixtures}
%
%
\date{\today}

\maketitle

\section{Dimensionless Equations}

In order to extract the essential parameters controlling the behaviour if the considered model, we non-dimensionalize it and introduce
\textit{characteristic units}.

\subsection{PDEs for chemotaxis}

Every physical quantity is written as a product of a non-dimensional quantity times a
normalization constant containing the units, e.g.
\begin{equation}
  \mathbf{x} = \tilde{\mathbf{x}}~x_u \; , \;\;
  t = \tilde{t}~t_u
\end{equation}
Here, all the quantities with a tilde are dimensionless while those
with a subscript " $_u$ " contain the units.
We also introduce the dimensionless chemical field,  $\tilde{c}^{p,h} = c^{p,h} x_u^d$,
where $d$ denotes the spatial dimension ($d=2$ in our case). Eq.~(2) in the main text then reads in dimensionless form as
\begin{equation}
  \partial_{\tilde{t}} \frac{\tilde{c}^{s}}{x_u^d} =
  \frac{D_c t_u}{x_u^2} \tilde{\Delta} \frac{\tilde{c}^{s}}{x_u^d}
  - k_d t_u \, \frac{\tilde{c}^{s}}{x_u^d}
  + \frac{k_0 t_u}{x_u^d} \sum_{i=1}^N\delta(\tilde{\mathbf{r}}-\tilde{\mathbf{r}}_i^{s})
  \; , \;\; s \in \{p,h\} \added{\;\; ,}
\end{equation}
where we have used
that $\delta(\alpha \mathbf{r}) = \frac{1}{\alpha^d} \delta(\mathbf{r})$.
Next, we choose $x_0 = R$, where $R$ denotes the (soft) radius of a particle and
$t_u = \frac{1}{k_0}$ yielding:
\begin{equation}
  \partial_{\tilde{t}} \tilde{c}^{s} = \tilde{D_c} \tilde{\Delta}
  \tilde{c}^{s} - \mu \tilde{c}^{s} +
  \sum_{i=1}^N\delta(\tilde{\mathbf{r}}-\tilde{\mathbf{r}}_i^{s}) \added{\;\; ,}
  \label{eq:chemotaxisPDEtilde}
\end{equation}
where the normalised coefficients $\tilde{D_c}$ and $\mu$
are then given by:  $\tilde{D_c} = \frac{D_c}{k_0 R^2}$ and $\mu = \frac{k_d}{k_0}$.

\subsection{Particle ODEs}
The same normalization is applied to Eq.~(1) in the main text which leads to:
\begin{eqnarray}
  \partial_{\tilde{t}} \tilde{\mathbf{r}}_i^{p} &=& \frac{t_u}{\gamma x_u^{d+2}} \left(\alpha_{pp} \tilde{\nabla} \tilde{c}^p(\tilde{\mathbf{r}}_i^{p},t)+\alpha_{ph} \tilde{\nabla} \tilde{c}^h(\tilde{\mathbf{r}}_i^{p},t)\right)
              - \tilde{\nabla} \tilde{V} + \sqrt{\frac{2 D t_u}{x_u^{2}}} \tilde{\bm{\eta}}_i^{p} \\
  \partial_{\tilde{t}} \tilde{\mathbf{r}}_i^{h} &=& \frac{t_u}{\gamma x_u^{d+2}} \left(\alpha_{hh} \tilde{\nabla} \tilde{c}^h(\tilde{\mathbf{r}}_i^{h},t)+\alpha_{hp} \tilde{\nabla} \tilde{c}^p(\tilde{\mathbf{r}}_i^{h},t)\right)
              - \tilde{\nabla} \tilde{V} + \sqrt{\frac{2 D t_u}{x_u^{2}}} \tilde{\bm{\eta}}_i^{h} ,
\end{eqnarray}
where $\bm{\eta}_i^{s}(t)$ represents unit-variance Gaussian white noise with zero mean.
With the additional definitions of $\tilde{\alpha}_{kl} = \frac{\alpha_{kl}}{\gamma k_0 R^{d+2}}$ and $\tilde{D} = \frac{D}{k_0 R^2}$ we obtain:

\begin{eqnarray}
  \partial_{\tilde{t}} \tilde{\mathbf{r}}_i^{p} &=& \tilde{\alpha}_{pp} \tilde{\nabla} \tilde{c}^p(\tilde{\mathbf{r}}_i^{p},t)+\tilde{\alpha}_{ph} \tilde{\nabla} \tilde{c}^h(\tilde{\mathbf{r}}_i^{p},t)
            -\tilde{\nabla} \tilde{V} + \sqrt{2 \tilde{D}} \tilde{\bm{\eta}}_i^{p} \label{eq:particlestilde1}\\
  \partial_{\tilde{t}} \tilde{\mathbf{r}}_i^{h} &=& \tilde{\alpha}_{hh} \tilde{\nabla} \tilde{c}^h(\tilde{\mathbf{r}}_i^{h},t)+\tilde{\alpha}_{hp} \tilde{\nabla} \tilde{c}^p(\tilde{\mathbf{r}}_i^{h},t)
            -\tilde{\nabla} \tilde{V} + \sqrt{2 \tilde{D}} \tilde{\bm{\eta}}_i^{h} . \label{eq:particlestilde2}
\end{eqnarray}

We model $V$ using the Weeks-Chandler-Anderson potential, with $\tilde{V}=\frac{1}{2} \sum_{i,j \neq i} \tilde{V}_{ij}$, where the sums run over all particles and where
\begin{eqnarray}
        \tilde{V}_{ij} = \left\{ \begin{array}{ll} 4 \tilde{\epsilon} \left[ (\frac{\tilde{\sigma}}{\tilde{r}_{ij}})^{12} - (\frac{\tilde{\sigma}}{\tilde{r}_{ij}})^6 \right] + \tilde{\epsilon} & \; \; $if$ \; \;  \tilde{r}_{ij} \leq 2^{1/6}\tilde{\sigma}  \\
                                          0                                                                                       & \; \; $if$ \; \;  \tilde{r}_{ij} > 2^{1/6}\tilde{\sigma}
                         \end{array} \right.
\end{eqnarray}

Here $r_{ij}$ denotes the distance between particle $i$ and $j$, $\tilde{r}_{ij}=r_{ij}/x_u$, $\tilde{\sigma}=\sigma/x_u$ and $\tilde{\epsilon}=\frac{\epsilon}{\gamma k_0 R^2}$.

\bigskip

In the following, we omit tildes. The specific value of $\epsilon$ hardly affects our results,
so the following independent and relevant dimensionless parameters remain:
\begin{enumerate}[label=\arabic*)]
  \item effective diffusion constants $D_c$ and $D$,
  \item reduced coupling constants $\alpha_{kl}$,
  \item ratio of decay and production of the chemical $\mu$
  \item density of species $p$,$h$ (which is conserved throughout the dynamics)
\end{enumerate}

\bigskip

\section{Linear Stability Analysis}

To understand the origin of the observed pattern formation, we perform a linear
stability analysis of Eqs.~(5,6) of the main text around the homogeneous solution
$(\rho^s,c^s)=(\rho_0^s,\rho_0^s/\mu)=:(\rho_0^s,c_0^s)$, $s \in \{p,h\}$.
The PDEs of the density in the continuous model for two different sorts
of particles $\rho^{s}=\rho^{s}(\mathbf{r},t)$, $s \in \{p,h\}$ read (in dimensionless units)
\begin{eqnarray}
  \partial_t \rho^{p} &=& D \Delta \rho^{p} - \alpha_{pp} \nabla \cdot (\rho^{p} \nabla c^{p}) - \alpha_{ph} \nabla \cdot (\rho^{p} \nabla c^{h})
  \label{eq:partial_rho1} \\
  \partial_t \rho^{h} &=& D \Delta \rho^{h} - \alpha_{hh} \nabla \cdot (\rho^{h} \nabla c^{h}) - \alpha_{hp} \nabla \cdot (\rho^{h} \nabla c^{p})    \; , \;\;
  \label{eq:partial_rho2}
\end{eqnarray}
where the corresponding PDEs for the phoretic field $c^{s}=c^{s}(\mathbf{r},t)$, $s \in \{p,h\}$
are given by:
\begin{eqnarray}
  \partial_t c^{s} = D_c \Delta c^{s} + \rho^{s} - \mu c^{s} \; . \;\;
  \label{eq:partial_c}
\end{eqnarray}

For the linear stability analysis, we consider small perturbations from the homogeneous steady state of the particle
density $\rho^{s}=\rho^{s}_0+ \delta\rho^{s}$ and the phoretic field
$c^{s}=c^{s}_0+ \delta c^{s}$.

The resulting linearised system of four equations describes
the time-evolution of the deviations of the particle density and the chemical field
from the homogeneous state $\rho^s=\rho_0^s$ and $c^s = \rho_0^s/\mu$.

\begin{eqnarray}
  \partial_t \delta\rho^{p} &=& D \Delta \delta\rho^{p} - \alpha_{pp} \rho^p_0 \Delta \delta c^{p} - \alpha_{ph} \rho^p_0 \Delta \delta c^{h}
  \label{eq:partial_rho1_taylor} \\
  \partial_t \delta\rho^{h} &=& D \Delta \delta\rho^{h} - \alpha_{hh} \rho^h_0 \Delta \delta c^{h} - \alpha_{hp} \rho^h_0 \Delta \delta c^{p}    \; , \;\;
  \label{eq:partial_rho2_taylor}
\end{eqnarray}
\begin{eqnarray}
  \partial_t  \delta c^{s} = D_c \Delta  \delta c^{s} +  \delta\rho^{s} - \mu \delta c^{s} \; , \;\;
  \label{eq:partial_c_taylor}
\end{eqnarray}

Since the linearised equations
Eqs.~(\ref{eq:partial_rho1_taylor}-\ref{eq:partial_c_taylor}) do not depend explicitly
on time, we perform a Fourier transformation in space
and make a separation Ansatz $\hat{\rho}^s(\mathbf{q},t)=e^{\lambda t} \hat{\rho^s}(\mathbf{q})$, $\hat{c}^s(\mathbf{q},t)=e^{\lambda t} \hat{c^s}(\mathbf{q})$, leading immediately to the following eigenvalue
problem (we set $\alpha_{hh}=0$ as in the main text):

\begin{eqnarray}
  \lambda
  \begin{pmatrix}
    \delta \hat{\rho}^{p} \\
    \delta \hat{c}^{p} \\
    \delta \hat{\rho}^{h} \\
    \delta \hat{c}^{h}
  \end{pmatrix}
  =
  \begin{pmatrix}
    -D \mathbf{q}^2 & \alpha_{pp} \rho^p_0 \mathbf{q}^2 & 0 & \alpha_{ph} \rho^{p}_0 \mathbf{q}^2 \\
    1 & -D_c \mathbf{q}^2 - \mu & 0 & 0 \\
    0 & \alpha_{hp} \rho^h_0 \mathbf{q}^2 & -D \mathbf{q}^2 & 0 \\
    0 & 0 & 1 & -D_c \mathbf{q}^2 -\mu
  \end{pmatrix}
  \begin{pmatrix}
    \delta \hat{\rho}^{p} \\
    \delta \hat{c}^{p} \\
    \delta \hat{\rho}^{h} \\
    \delta \hat{c}^{h}
  \end{pmatrix} \;\; ,
\end{eqnarray}
where the stability of the system is determined by the eigenvalues $\lambda$ of the
matrix, as follows:
\begin{enumerate}[label=(\alph*)]
  \item The steady state is stable, if the eigenvalues of the matrix all have real
  parts strictly less than zero.
  \item The steady state is unstable, if at least one of the eigenvalues of the
  matrix has a positive real part.
  \item Otherwise in the marginal case higher order terms determine the stability of the problem.
\end{enumerate}

The four eigenvalues $\lambda$ of the linearised system are  explicitly given by ($\alpha_{hh}=0$, $\rho^p_0=\rho^h_0=\rho_0$):
\begin{equation}
  \lambda=-\frac{q^2}{2}(D_c+D+\frac{\mu}{q^2}) \pm \frac{1}{2}\sqrt{[q^2(D_c+D+\frac{\mu}{q^2})]^2-4(D_c+\frac{\mu}{q^2})Dq^4+2 \alpha_{pp} \rho_0 q^2 \pm 2 q^2 \rho_0 \sqrt{4\alpha_{ph} \alpha_{hp}+\alpha_{pp}^2}} \; ,
  \label{eq:eigenvalues}
\end{equation}
%

For the parameter range under consideration, eigenvalues with non-vanishing imaginary part exist ($\mathrm{Im}(\lambda) \neq
0)$, if the condition $-4\alpha_{ph} \alpha_{hp} > \alpha_{pp}^2$ is fulfilled.
In order to develop an instability for long wavelengths  ($|q| \longrightarrow 0$), a Taylor series extension of (\ref{eq:eigenvalues}) and an examination where $\mathrm{Re}(\lambda)>0$ provides us with the criterion
\begin{equation}
  2D \mu < \alpha_{pp} \rho_0.
\end{equation}
If the eigenvalues are real (here, $\mathrm{Im}(\lambda) = 0 \iff -4\alpha_{ph}
\alpha_{hp} < \alpha_{pp}^2$), then we obtain the following criterion for instability from calculating $\mathrm{Re}(\lambda)>0$:
\begin{equation}
  2D \mu < \rho_0 \left(\alpha_{pp}+\sqrt{4\alpha_{ph} \alpha_{hp} + \alpha_{pp}^2}\right)
\end{equation}
Altogether, an instability is given, when the following criterion is
fulfilled:
\begin{equation}
  2D \mu < \rho_0 \mathrm{Re} \left[ \alpha_{pp}+\sqrt{4\alpha_{ph} \alpha_{hp} + \alpha_{pp}^2} \right]
\end{equation}
and this instability is also oscillatory if
\begin{equation}
  -4\alpha_{ph} \alpha_{hp} > \alpha_{pp}^2 \; .
\end{equation}

\bigskip

\section{Description of numerical methods}

\subsection{Numerical methods for solving the equations of the particle-based chemotaxis model}

We solve our model Eqs.~(\ref{eq:chemotaxisPDEtilde},\ref{eq:particlestilde1},\ref{eq:particlestilde2})
by discretizing space and time. We perform
simulations in two spatial dimensions (2D) with periodic boundary conditions.
Interactions between the particles are described using the
Weeks-Chandler-Anderson potential.
The PDEs (\ref{eq:chemotaxisPDEtilde}) for the dynamics of the chemical fields are
solved in time with a forward Euler method. This does not impose any restriction
on the timestep $dt$ due to the Courant-Friedrich-Lewy (CFL) condition
\cite{courant-friedrichs-lewy:1928} of the heat equation
($D_c\left(\frac{dt}{dx^2}+\frac{dt}{dy^2}\right) < \frac{1}{4}$, where $dt$ and
$dx$, $dy$ denote the temporal and spatial discretisations), since the necessary
timestep $dt$ for the particle dynamics is small enough to not violate the above
CFL condition. The Laplace operator of the heat term is approximated with
central differences. The particle dynamics of the colloids
(Eqs.~\ref{eq:particlestilde1},\ref{eq:particlestilde2}) is solved in time with the Euler-Maruyama scheme
\cite{higham:2001} for incorporation the noise term. The interactions of the
colloids is effectively treated by applying a cell-list summation
\cite{allen-tildesley:1989}.

In order to speed up the calculations, we make use of cell lists
\cite{allen-tildesley:1989}, with a minimal image convention. This is motivated
by the fact that the repulsive force caused by the potential $V$ is short ranged
and hence we can introduce a cutoff radius $r_c=2^{1/6} \sigma$. More explicitly, we
consider only interactions within a cutoff radius $r_c$ such that the pair
interaction $V_{ij}=0$ for $r_{ij}>r_c$.

\bigskip

Different simulations are performed by varying the diffusion constants $D$ and $D_c$,
the coupling constants $\alpha_{kl}$ and the ratio of decay and production of
the chemical $\mu$.

\clearpage

\begin{samepage}
\begin{center}
\textit{Typical parameters used in the simulation of the dimensionless Eqs.~(\ref{eq:chemotaxisPDEtilde},\ref{eq:particlestilde1},\ref{eq:particlestilde2}):}

\bigskip

\begin{tabular}{lll}
  $\#$ of particles $N$               & \phantom{bla} &$2*10^3$\\
  WCA-potential amplitude $\epsilon$  & &1 \\
  coupling constants                  & &$-10 < \alpha_{kl} < 10$ \\
  diffusion constant $D_c$            & &$1$ \\
  diffusion constant $D$              & &$0.001 - 1$ \\
  decay/production of $c$             & &$0.001 < \mu < 1 $ \\
  time step                           & &$\Delta t = 0.001 t_0$ \\
  simulation time                     & &$\sim 10^4 t_0$ \\
  grid for chemical                   & &$250$ x $250$ grid cells
\end{tabular}
\end{center}
\end{samepage}

\bigskip

\subsection{Numerical methods for solving the equations of the continuum chemotaxis model}

In addition to the particle-based simulations, we have also numerically solved
the system of the four coupled dimensionless PDEs of the continuum model,
Eqs.~(\ref{eq:partial_rho1},\ref{eq:partial_rho2},\ref{eq:partial_c}) as before
in two spatial dimensions with periodic boundary conditions. We use a forward
Euler method to propagate the densities $\rho^{s}$ and the chemical fields
$c^{s}$ in time. The Laplace operator is approximated with central differences.

\subsubsection{Repulsion between cells}

We must be careful when implementing the chemotactic continuum model, which unlike
our particle based equations, neglect the steric repulsions among the particles,
preventing sharp density peaks, destabilizing the simulation.
In view of this effect, we add an
additional nonlinear term to the continuum model to effectively describe local repulsions among the particles \cite{Tumor-Macrophage}:
\begin{eqnarray}
  \partial_t \rho^{s} &=& D \Delta \rho^{s} - \sum_{s \in \{p,h\}} \alpha_{s s'} \nabla \cdot (\rho^{s} \nabla c^{s'}) + D_{\text{rep}}^s \nabla \cdot (\rho^{s} \nabla (\rho^{p} + \rho^{h})) \, .
  \label{eq:partial_rho_reg1}
\end{eqnarray}
Here, $D_{\text{rep}}^s$ with $s \in \{p,h\}$ are dimensionless repulsion coefficients and we assume that
the strength of the repulsion is the same between the two different species,
$D_{\text{rep}}^p=D_{\text{rep}}^h$.

\subsubsection{Regularization of the density}

As a second adjustment of the continuum model, avoiding a blow up
\cite{kurganov_medvidova:2014} we use:
\begin{eqnarray}
  \partial_t \rho^{s} &=& D \Delta \rho^{s} - \sum_{s \in \{p,h\}} \alpha_{s s'} \nabla \cdot \left( \frac{\rho^{s}}{1 + \kappa \rho^{s}} \nabla c^{s'} \right) + D_{\text{rep}}^s \nabla \cdot (\rho^{s} \nabla (\rho^{p} + \rho^{h})) \, ,
  \label{eq:partial_rho_reg2}
\end{eqnarray}
where $\kappa$ is a small regularization parameter and $\kappa \rightarrow 0$
leads to the original system. In \cite{kurganov_medvidova:2014} it was shown
that the solutions of equations of the kind of Eqs.~(\ref{eq:partial_rho1},\ref{eq:partial_rho2}) typically have a spiky
structure and Eq.~(\ref{eq:partial_rho_reg2}) can be used to model the aggregation
phenomena.

\section{Mean particle velocity}
In order to distinguish between the different phases, in particular between the
clusters and the hunting swarm phase, we calculate the mean particle velocity, which is
defined as
\begin{equation}
  v(t)=\left\langle \frac{1}{2N} \sum_{\substack{i=1,N \\ s=p,h}} \left| \frac{\mathbf{r}_i^s(t+dt)-\mathbf{r}_i^s(t)}{dt} \right| \right\rangle
\end{equation}
where we average over all particles of species $s \in \{p,h\}$ and $\left\langle \cdot \right\rangle$ denotes the ensemble average.

\section{Diffusion equation in 3D}
\begin{figure}[H]
  \begin{subfigure}[b]{0.5\textwidth}
    \centerline{\includegraphics[width=0.7\textwidth]{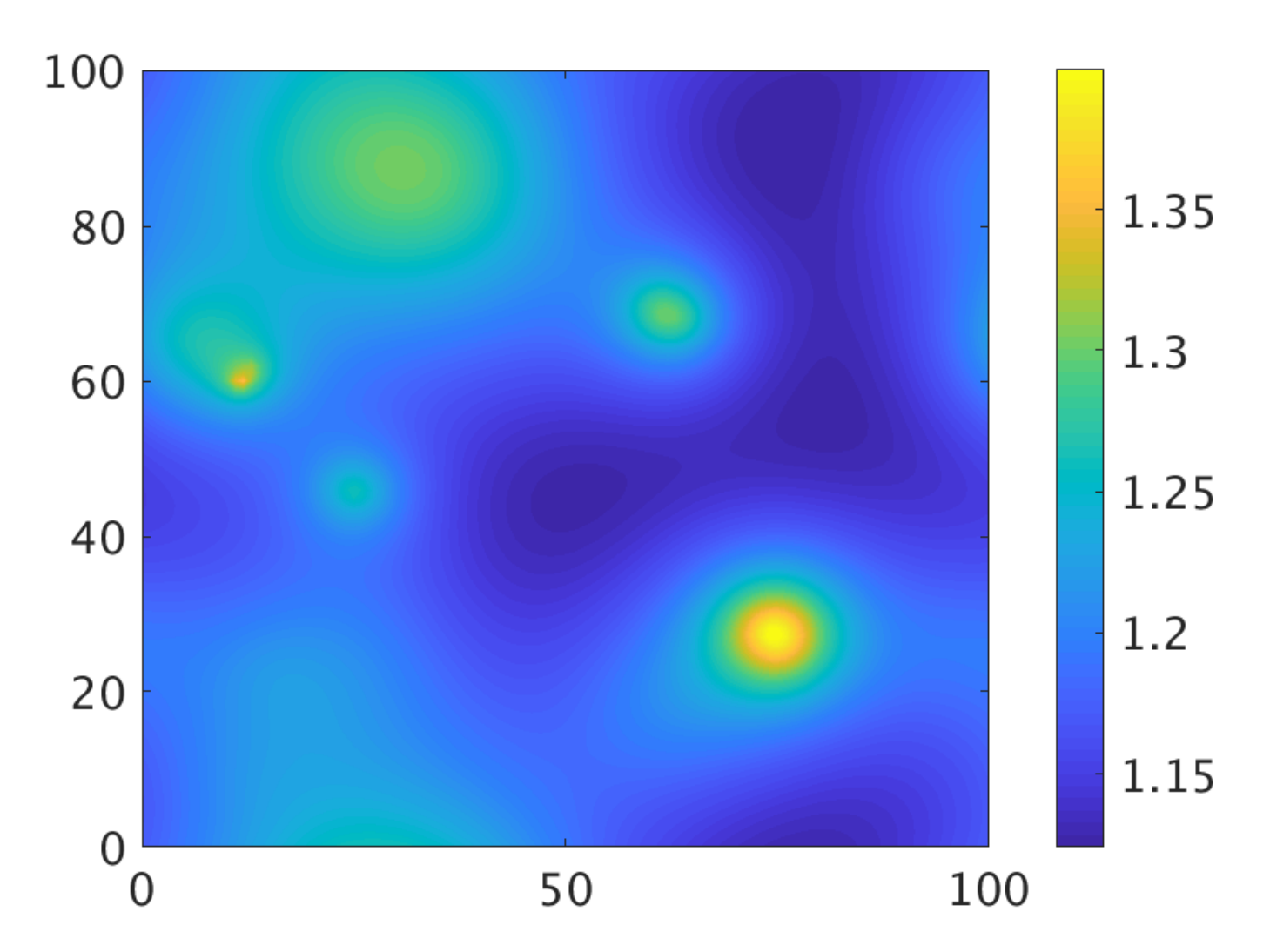}}
  \end{subfigure}
  \begin{subfigure}[b]{0.5\textwidth}
      \centerline{\includegraphics[width=0.8\textwidth]{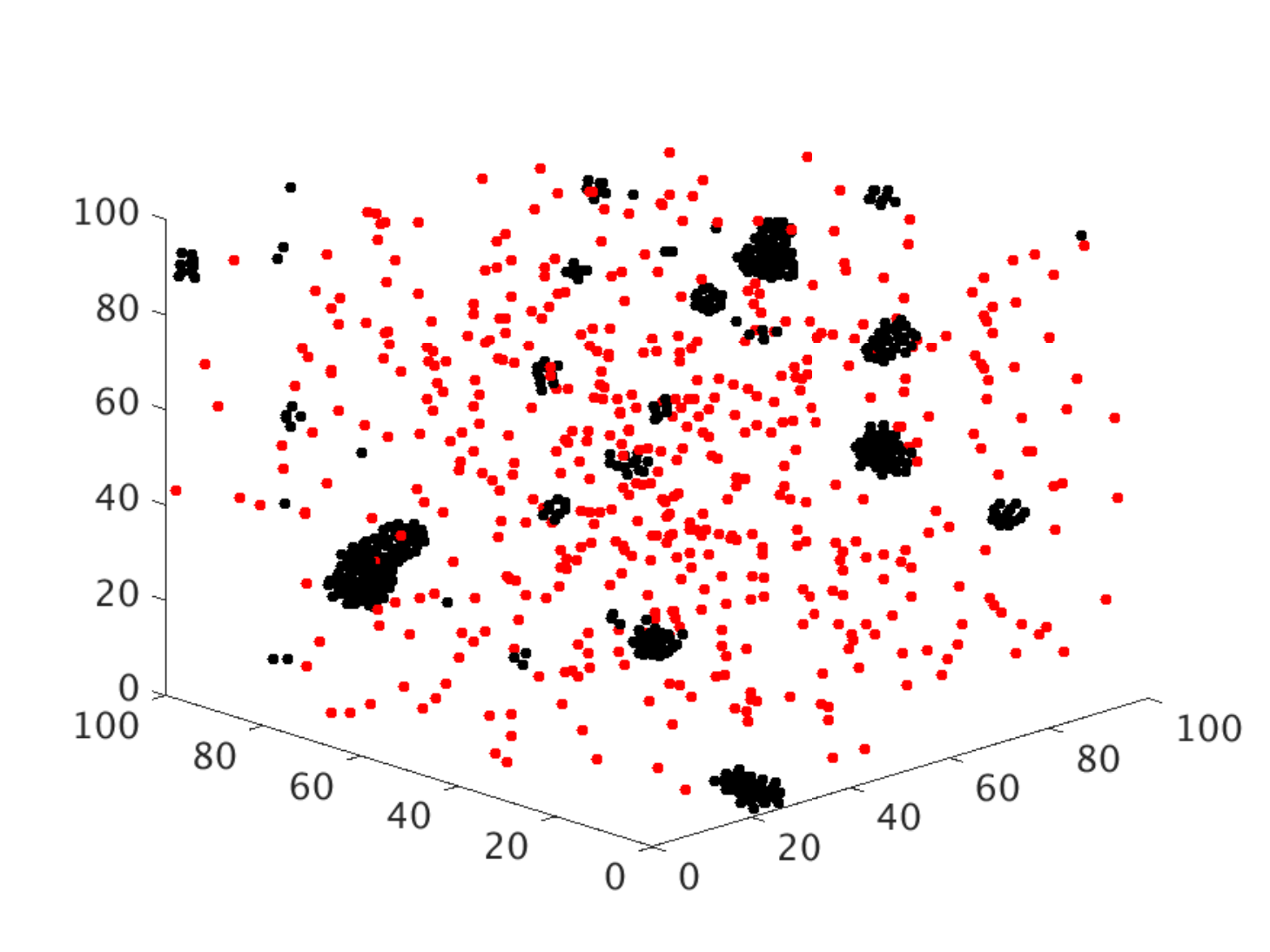}}
  \end{subfigure}
  \caption{
  \label{fig:snapshot_3D}
  Simulation snapshot of the particle based model in 3D. Left panel shows the diffusion of $c^p$ by looking at the cross section of the chemical field in a plane perpendicular to the $z$-axis. Right panel shows the prey and the hunters, colored in black and red, respectively. Parameters as in Fig.~2(b) of the main text.
  }
\end{figure}

To verify that our results do not change qualitatively when solving the diffusion equation in 3D,
we have performed additional simulations for verification. Figure \ref{fig:snapshot_3D} shows an
exemplaric simulation snapshot of the particle based model in three dimensions for a cluster in the blue domain ($\alpha_{hp}=0.01,\alpha_{ph}=-10$).

\section{Description of the movies}

The movies 1-8 show the time-evolution of the different patterns, hunting swarms,
mixed clusters and core-shell clusters, for the particle-based and the continuum
model for the same parameters as used in Fig.~2 of the main text (panels (a-h)
corresponding to the files mov1-mov8).
Movie 9 shows a dynamical inside out reversal for an
appropriate initial condition, starting with a dense and comparatively large cluster of prey particles, which is
then invaded by the hunter particles. Parameters used for movie 9:
$\alpha_{pp}=1$, $\alpha_{hh}=0$, $\alpha_{hp}=10$, $\alpha_{ph}=-0.01$, $\mu=0.001$, $D_c=0.1$, $D=0.003$, $\epsilon=1$, $2N=300$ particles and $L_{\text{box}}=80$.

\newpage
\section*{References}

\bibliography{paper2}